\documentclass{article}

\usepackage{amsmath,amssymb,amsthm,amscd,euscript,amsfonts}
\usepackage{mathtools}
\usepackage{color}
\usepackage[pdftex]{graphicx}
\usepackage[toc,page]{appendix}
\usepackage[utf8]{inputenc}
\usepackage[nottoc]{tocbibind}
\usepackage{authblk}
\usepackage{cite}
\DeclareGraphicsExtensions{pdf, jpg, jpeg, png}

\renewcommand{\o}{\otimes}
\newcommand{\eps}{\varepsilon}

\input xy
\usepackage[color,matrix,arrow]{xy}
\xyoption{all}

\def \c{\mathbb C}
\def \r{\mathbb R}
\def \l{\mathcal L}
\def \h{\mathcal H}
\def \k{\mathbf k}

\newtheorem{theorem}{Theorem}[section]

\newtheorem{remark}[theorem]{Remark}

\title{An integrable case of the $p+ip$ pairing Hamiltonian \\
interacting with its environment}
\author{Inna Lukyanenko, Phillip S. Isaac, Jon Links}
\date{}

\begin{document}
\input epsf

\maketitle

\vspace{-1.1cm}
\begin{abstract}
We consider a generalisation of the $p+ip$ pairing Hamiltonian with external interaction terms. These terms allow for the exchange of particles between the system and its environment. As a result the $\mathfrak{u}(1)$ symmetry associated with conservation of particle number, present in the $p+ip$ Hamiltonian, is broken.  Nonetheless the generalised model is integrable. We establish integrability using the Boundary Quantum Inverse Scattering Method, with one of the reflection
matrices chosen to be non-diagonal. We also derive the corresponding Bethe Ansatz Equations, the roots of which parametrise the exact
solution for the energy spectrum.
\end{abstract}
\vspace{-0.4cm}
\tableofcontents


\section{Introduction}

Understanding how quantum systems interact with their environment, and being able to control such interactions, is a major challenge facing quantum engineering. One such framework where this applies is provided by Josephson junctions, fabricated through weakly-coupled superconductors. These structures have received widespread study as a potential architecture for the coherent control of quantum bits e.g. \cite{mss01,yn05,cw08,t10,petal}. In some instances, such as the Cooper-pair box Josephson junction, the system is described in terms of a simple Bose--Hubbard tunneling model \cite{mss01,yn05}. More refined analyses to produce insights into environment interactions have also been undertaken, in particular through explicit use of the degrees of freedom of the $s$-wave pairing Hamiltonian in the strong-coupling limit \cite{a10,am13}.  

In recent years the $p+ip$ pairing Hamiltonian has emerged as an example of a superconducting model which is integrable, and admits an exact Bethe Ansatz solution \cite{ilsz09}. The solution was obtained by application of the Quantum Inverse Scattering Method (QISM)
\cite{tf79,ks79} associated with the {\it trigonometric} $XXZ$ solution of the Yang--Baxter equation in the quasi-classical limit. This result has led to several studies of the model, including re-derivations of the solution from different perspectives such as the Richardson-Gaudin approach, investigations into the ground-state structure,  and extending the application of the exact solution for the calculation of correlation functions \cite{s09,dilsz10,rdo10,ris14,vdv14,cdvv15,lmm15}.   

Here we will establish that there is an extension of the $p+ip$ pairing model, involving interaction terms coupling to the environment, which maintains integrability. These interaction terms allow for the exchange of particles between the system and its environment. As such they break the $\mathfrak{u}(1)$ invariance associated with conservation of particle number which is present in the $p+ip$ Hamiltonian. From a 
na\"{\i}ve perspective it appears that the prospect for constructing an integrable extension of the $p+ip$ pairing Hamiltonian with broken $\mathfrak{u}(1)$ symmetry is dire. The $p+ip$ pairing Hamiltonian is constructed as a linear combination of mutually conserved operators with co-efficients which are {\it dependent} on the particle number \cite{s09,dilsz10,rdo10,vdv14}. If the particle number is not conserved, this approach fails. Fortuitously, our recent study \cite{lil14} uncovered a means to overcome this issue. It turns out that the $p+ip$ pairing Hamiltonian and its exact solution can also be derived through  application  of the {\it Boundary} Quantum Inverse Scattering Method (BQISM) \cite{skl88} associated with the {\it rational} $XXX$ solution of the Yang-Baxter equation in the quasi-classical limit. In this approach, the    
Hamiltonian is constructed as a linear combination of mutually conserved operators with co-efficients which are {\it independent} of the particle number. Thus this framework can be extended to produce a generalised model interacting with the environment, as we will describe below. 

In Sect. 2 we introduce the Hamiltonian and review the Boundary Quantum Inverse Scattering Method of Sklyanin \cite{skl88}. Sect. 3 details a construction to obtain a set of mutually commuting operators, and shows how the Hamiltonian may be expressed in terms of the elements of this set. Sect 4. then utilises Bethe Ansatz results to obtain the exact solution for the Hamiltonian. Concluding remarks are given in Sect. 5  


\section{Preliminaries}

\subsection{The pairing model interacting with its environment}

We first introduce the isolated pairing mode not interacting with the environment. Let $c_\k,
c_\k^\dagger$ denote the annihilation and creation operators for two-dimensional fermions of mass
$m$ and momentum $\k=(k_x,k_y)$. Then the pairing Hamiltonian is 
\begin{equation*}
H_0=\sum_{\k}\frac{|\k|^2}{2m} c_\k^\dagger c_\k
-\frac{G}{4}\sum_{\k\neq\pm\k^\prime}(k_x+i k_y)(k_x^\prime-i k_y^\prime)c_\k^\dagger c_{-\k}^\dagger c_{-\k^\prime}c_{\k^\prime},
\end{equation*}
where $G\in\r$ is a constant and the summation is taken over all momentum states $\k$. 
The annihilation and creation operators $c_\k, c_\k^\dagger$ satisfy the canonical anticommutation relations:
\begin{equation*}
\{c_\k,c_{\k^\prime}\}=\{c_\k^\dagger,c_{\k^\prime}^\dagger\}=0,\ \{c_\k,c_{\k^\prime}^\dagger\}=\delta_{\k\k^\prime}I.
\end{equation*}

Now consider a more general Hamiltonian with an extra term
\begin{equation}
H=H_0+\frac{\Gamma}{2}\sum_{\k}\left((k_x+ik_y)c_\k^\dagger c_{-\k}^\dagger+(k_x-ik_y)c_{-\k}c_\k\right),
\label{env}
\end{equation}
where $\Gamma\in\r$ is a constant. We note that this Hamiltonian is Hermitian, and the extra
term can be interpreted as creation and annihilation of pairs of fermions, resulting from 
interaction with the environment. It is important to distinguish this type of interaction with the environment from other examples, e.g. \cite{bm78} in the context of a heat bath, which facilitate a notion of entanglement with the environment. In our model there is no entanglement between the system and the environment, because the state space for the environment is not explicitly defined. We will comment further on this aspect in the Conclusion.  

We now restrict to the Hilbert subspace that allows only paired particle states. By
imposing this restriction, we do not consider states on which the operators in the interaction term in the
Hamiltonian (i.e. the second term) has trivial action. 
On this subspace the following equality is satisfied:
\begin{equation}\label{firstequ}
2c_\k^\dagger c_\k c_{-\k}^\dagger c_{-\k}=c_\k^\dagger c_\k+c_{-\k}^\dagger c_{-\k}.
\end{equation}
Set $z_\k=|\k|$ and $k_x+i k_y=|\k|\text{exp}(i\phi_\k)$. Introduce the following notation:
\begin{equation*}
S_\k^+=\text{exp}(i\phi_\k)c_\k^\dagger c_{-\k}^\dagger,\ S_\k^-=\text{exp}(-i\phi_\k)c_{-\k} c_\k,\ 
S_\k^z=c_\k^\dagger c_{-\k}^\dagger c_{-\k} c_\k-\frac{I}{2}.
\end{equation*}
\begin{remark}
On this restricted subspace, one may verify the $\mathfrak{su}(2)$ algebra commutation relations:
\begin{equation*}
[S_\k^z,S_\k^{\pm}]=\pm S_\k^{\pm},\ [S_\k^+,S_\k^-]=2S_\k^z.
\end{equation*}
\end{remark}
We now use integers to enumerate the unblocked pairs of momentum states ($\k$ and $-\k$). Working
in units such that $m=1$, using equation (\ref{firstequ}) and ignoring the constant term
$\displaystyle{\frac{1}{2}\sum_{k=1}^\l z_k^2},$ we obtain
\begin{equation*}
H_0=\sum_{k=1}^\l z_k^2 S_k^z-G\sum_{k=1}^\l\sum_{j\neq k}z_k z_j S_k^+ S_j^-,
\end{equation*}
which exhibits $\mathfrak{u}(1)$-symmetry associated with the operator
$\displaystyle{S^z=\sum_{k=1}^\l S_k^z}.$
The full Hamiltonian that we work with therefore becomes
\begin{equation}\label{H}
H=H_0+\Gamma\sum_{k=1}^\l z_k\left(S_k^++S_k^-\right).
\end{equation}
This Hamiltonian no longer possesses $\mathfrak{u}(1)$-symmetry. 

In this article we show that the Hamiltonian (\ref{H}) is integrable by means of the BQISM. 
Recently, a systematic method, referred to as the \textit{Off-Diagonal Bethe Ansatz}
(ODBA) has been proposed for solving such models \cite{cysw13}. This method has been since applied
to several long-standing problems and the results has been summarised in the recent book by Wang et al.
\cite{wycs15}. Based on the results from \cite{hcyy15}, we derive
the formulae for the eigenvalues of the conserved operators, the corresponding Bethe Ansatz Equations (BAE) and the energy (i.e. the eigenvalue of the Hamiltonian). 


\subsection{Boundary Quantum Inverse Scattering Method}

In this section we review the Sklyanin's BQISM
\cite{skl88} and specify the ingredients in the context of our model. 
Throughout this paper we fix a vector space $V=\c^2$. 
A key element of the BQISM is the $R$-matrix, which is an invertible operator 
$R(u)\in\text{End}(V\o V)$ depending on a spectral parameter $u\in\c$ and satisfying
the \textit{Yang-Baxter Equation} (YBE) in $\text{End}(V\o V\o V)$
\begin{equation*}
R_{12}(u-v)R_{13}(u)R_{23}(v)=R_{23}(v)R_{13}(u)R_{12}(u-v).
\end{equation*}
Here, as usual, the subscripts indicate the spaces in which the corresponding $R$-matrix acts non-trivially.

In this paper we consider the rational $R$-matrix that is usually associated with the XXX spin chain
\begin{equation*}
R(u)=uI+\eta P=
                   \begin{pmatrix}
                    u+\eta & 0 & 0 & 0\\
                    0 & u & \eta & 0\\
                    0 & \eta & u & 0\\
                    0 & 0 & 0 & u+\eta\\
                    \end{pmatrix},
\end{equation*}
where $\eta\in\c$ is the quasi-classical parameter and $P$ is the permutation operator in $V\o V$.

In the BQISM framework the boundary conditions are encoded in the left and right \textit{reflection
matrices}, or \textit{$K$-matrices}, $K^-(u)$ and $K^+(u)\in\text{End}(V)$, which satisfy the
\textit{reflection equations} in $\text{End}(V\o V)$ 
\begin{align}
R_{12}(u-v)K_1^-(u)R_{21}(u+v)K_2^-(v)&=K_2^-(v)R_{12}(u+v)K_1^-(u)R_{21}(u-v),\label{reflection-}\\
R_{12}(v-u)K_1^+(u)R_{21}(-u-v-2\eta)K_2^+(v)&=K_2^+(v)R_{12}(-u-v-2\eta)K_1^+(u)R_{21}(v-u) \label{reflection+}.
\end{align}
One can check that the following $K$-matrix satisfies the first reflection equation (\ref{reflection-}):
\begin{equation*}
K^-(u)=\begin{pmatrix}
                 \xi^-+u & \psi^- u\\
                 \phi^- u & \xi^--u
                 \end{pmatrix}.
\end{equation*}
Then, 
\begin{equation*}
K^+(u)=-K^-(-u-\eta)|_{\xi^-\mapsto-\xi^+,\psi^-\mapsto\psi^+,\phi^-\mapsto\phi^+} = 
                 \begin{pmatrix}
                 \xi^++u+\eta & \psi^+(u+\eta)\\
                 \phi^+(u+\eta) & \xi^+-u-\eta
                 \end{pmatrix}
\end{equation*}
automatically satisfies the dual reflection equation (\ref{reflection+}).

We may express the Hilbert space of states in the form
\begin{equation}\label{hilbert}
\h=\bigotimes_{j=1}^{\l}V_j=V^{\o\l},
\end{equation}
where each local space $V_j$  (a copy of $V$) is a fixed representation space for the
$\mathfrak{su}(2)$ algebra spanned by $S_j^-,S_j^+,S_j^z$ (indices indicate in which space the
corresponding operator acts non-trivially). For each label $j$ in the tensor product (\ref{hilbert}),
we introduce the \textit{Lax operator}
\begin{equation}\label{L-spin-rat}
L_{aj}(u)=\frac{1}{u}\begin{pmatrix}
                                     u+\eta S_j^z & \eta S_j^-\\
                                    \eta S_j^+ & u-\eta S_j^z
                                     \end{pmatrix}=I+\frac{\eta}{u}\begin{pmatrix}
                                                                                         S_j^z & S_j^-\\
                                                                                         S_j^+ & -S_j^z
                                                                                         \end{pmatrix}\in\text{End}(V_a\o V_j),
\end{equation}
where the auxiliary space $V_a$ is another copy of $V$. 

It is straightforward to check that the Lax operator (\ref{L-spin-rat}) satisfies the \textit{RLL relation} in $\text{End}(V_a\o V_b\o V_j)$
\begin{equation}\label{RLL}
R_{ab}(u-v)L_{aj}(u)L_{bj}(v)=L_{bj}(v)L_{aj}(u)R_{ab}(u-v),
\end{equation}
where $V_{b}=V$ is another auxiliary space.
\begin{remark}
Note that the Lax operator (\ref{L-spin-rat}) satisfies the following property:
\begin{equation}\label{L-inverse}
L_{aj}(u)L_{aj}(\eta-u)=\left(1+\eta^2\frac{s_j(s_j+1)}{u(\eta-u)}\right)I,
\end{equation}
where $s_j$ is the value of the spin on the local space $V_j$.
\end{remark}
Define the \textit{monodromy matrix} as
\begin{equation}\label{T}
T_a(u)=L_{a\l}(u-\eps_\l)\ldots L_{a1}(u-\eps_1),
\end{equation}
where $\eps_j\in\c$ are the inhomogeneity parameters.
From the RLL relation (\ref{RLL}) it follows that the monodromy matrix (\ref{T}) satisfies the
\textit{RTT relation} in $\text{End}(V_a\o V_b\o\h)$
\begin{equation}\label{RTT}
R_{ab}(u-v)T_a(u)T_b(v)=T_b(v)T_a(u)R_{ab}(u-v).
\end{equation}
Let us construct the \textit{dual monodromy matrix} as
\begin{equation*}
\tilde{T}_a(u)=L_{a1}(u+\eps_1+\eta)\ldots L_{a\l}(u+\eps_{\l}+\eta).
\end{equation*}
From the property (\ref{L-inverse}) of the Lax operator it follows that
\begin{equation*}
\tilde{T}_a(u)\propto L_{a1}^{-1}(-u-\eps_1)\ldots L_{a\l}^{-1}(-u-\eps_{\l})=T^{-1}_a(-u),
\end{equation*}
which implies that $\tilde{T}_a(u)$ satisfies the following relations:
\begin{align}
\tilde{T}_b(v)R_{ab}(u+v)T_a(u)=T_a(u)R_{ab}(u+v)\tilde{T}_b(v),\label{RTT'}\\
\tilde{T}_a(u)\tilde{T}_b(v)R_{ab}(v-u)=R_{ab}(v-u)\tilde{T}_b(v)\tilde{T}_a(u). \label{RTT''}
\end{align}
Now, the \textit{double row monodromy matrix} is constructed as follows:
\begin{equation*}
\mathcal{T}_a(u)=T_a(u)K_a^-(u)\tilde{T}_a(u),
\end{equation*}
and the relations (\ref{RTT}), (\ref{RTT'}) and (\ref{RTT''}) imply that it satisfies 
\begin{equation}\label{RTRT}
R_{ab}(u-v)\mathcal{T}_a(u)R_{ba}(u+v)\mathcal{T}_b(v)=\mathcal{T}_b(v)R_{ba}(u+v)\mathcal{T}_a(u)R_{ab}(u-v).
\end{equation}
The \textit{double row transfer matrix} is then defined as  
\begin{equation}\label{t-b0}
t(u)={\rm tr}_a\left(K^+_a(u)\mathcal{T}_a(u)\right).
\end{equation}
Using (\ref{RTRT}), one can show that these transfer matrices (\ref{t-b0}) commute for any two
values of the spectral parameter:
\begin{equation*}
[t(u),t(v)]=0 \text{\ \ for all\ \ }u,v\in\c.
\end{equation*} 
Thus, (\ref{t-b0}) can be used as a generating function for the conserved operators of the system.

In what follows, it is convenient to make a variable change $u\mapsto u-\eta/2,\
\eps_j\mapsto\eps_j-\eta/2$ and redefine all functions taking this into account. This results in
\begin{align}
K^-(u)=&\begin{pmatrix}
                 \xi^- +u-\eta/2 & \psi^- (u-\eta/2)\\
                 \phi^- (u-\eta/2) & \xi^- -u+\eta/2
                 \end{pmatrix}, \label{K-}\\
K^+(u)=&\begin{pmatrix}
                 \xi^++u+\eta/2 & \psi^+(u+\eta/2)\\
                 \phi^+(u+\eta/2) & \xi^+-u-\eta/2
                 \end{pmatrix}, \label{K+}
\end{align}
\begin{equation*}
\tilde{T}_a(u)=L_{a1}(u+\eps_1)\ldots L_{a\l}(u+\eps_{\l}).
\end{equation*}
Thus, the transfer matrix (\ref{t-b0}) will take the following form:
\begin{equation}\label{t-b}
t(u)={\rm tr}_a\left(K_a^+(u)L_{a\l}(u-\eps_{\l})\ldots  L_{a1}(u-\eps_1)K_a^-(u)L_{a1}(u+\eps_1)...L_{a\l}(u+\eps_{\l})\right).
\end{equation}


\section{Construction of the conserved operators and Hamiltonian}

As discussed in the introduction, our focus will be on taking the quasi-classical limit
$\eta\rightarrow 0$, thus connecting our study to the Richardson-Gaudin class of models
\cite{rdo10,vdv14,cdvv15,hcyy15,cms14,cmrs15}. Indeed, the resulting model we refer to as the {\it open, rational
Richardson-Gaudin model}. 
To be able to take this quasi-classical limit, however, we require that the $K$-matrices satisfy the following condition:
\begin{equation}\label{condition}
K^+(u)K^-(u)\rightarrow f(u)I \ \ \text{as}\ \ \eta\rightarrow 0.
\end{equation}
Assume the following dependence of the parameters on $\eta$:
\begin{equation}\label{eta}
\begin{aligned}
\xi^+&=\xi+\eta\alpha, \ \ \psi^+=\psi+\eta\gamma, \ \ \phi^+=\phi+\eta\lambda,\\
\xi^-&=-\xi+\eta\beta, \ \ \psi^-=\psi+\eta\delta, \ \ \phi^-=\phi+\eta\mu.\\
\end{aligned}
\end{equation}
Now consider
\begin{equation*}
K^+(u)K^-(u)|_{\eta=0}=\begin{pmatrix}
                                            \xi+u & \psi u\\
                                            \phi u & \xi-u
                                            \end{pmatrix}\begin{pmatrix}
                                                                   -\xi +u & \psi u\\
                                                                   \phi u & -\xi-u
                                                                   \end{pmatrix}=(u^2(1+\psi\phi)-\xi^2)I.
\end{equation*}
Thus, the condition (\ref{condition}) is satisfied. Now, expanding the $K$-matrices in $\eta$ we obtain
\begin{equation}\label{K-expansion}
K^+(u)=K^+_1(u)+\eta K^{+}_2(u)+o(\eta)
\end{equation}      
with
\begin{equation*}
K^+_1(u)=\begin{pmatrix}
                   \xi+u & \psi u\\
                   \phi u & \xi-u
                   \end{pmatrix}, \
K^+_2(u)=\begin{pmatrix}
                   \alpha+1/2 & \gamma u+\psi/2\\
                   \lambda u+\phi/2 & \alpha-1/2
                   \end{pmatrix},
\end{equation*}
and
\begin{equation}\label{K+expansion}
K^-(u)=K^-_1(u)+\eta K^-_2(u)+o(\eta)
\end{equation}      
with
\begin{equation*}
K^-_1(u)=\begin{pmatrix}
                   -\xi+u & \psi u\\
                   \phi u & -\xi-u
                   \end{pmatrix}, \
K^-_2(u)=\begin{pmatrix}
                   \beta-1/2 & \delta u-\psi/2\\
                   \mu u-\phi/2 & \beta+1/2
                   \end{pmatrix}.
\end{equation*}
For the Lax operator we have
\begin{equation}\label{L-expansion}
L_{aj}(u)=I+\frac{\eta}{u}\ell_{aj}, \ \text{with}\ \ell_{aj}=\begin{pmatrix}
                                                                                                        S^z_j & S^-_j\\
                                                                                                        S^+_j & -S^z_j
                                                                                                       \end{pmatrix}.
\end{equation}


\subsection{The first family of conserved operators}\label{sec:im}

In the quasi-classical limit, the conserved operators $\tau_j$ are constructed as follows from the transfer matrix (\ref{t-b}):
\begin{equation*}
\lim_{u\rightarrow\eps_j}(u-\eps_j)t(u)=\eta^2\tau_j+o(\eta^2).
\end{equation*}
Substituting (\ref{K-expansion}), (\ref{K+expansion}) and (\ref{L-expansion}) into (\ref{t-b}) we obtain
\begin{equation*}
\begin{aligned}
\lim_{u\rightarrow\eps_j}(u-\eps_j)t(u)
=&\,\eta{\rm tr}_a\Bigg[K^+_{1a}(\eps_j)\ell_{aj}K^-_{1a}(\eps_j)+\eta\sum_{k>j}^{\l}\frac{K^+_{1a}(\eps_j)\ell_{ak}\ell_{aj}K^-_{1a}(\eps_j)}{\eps_j-\eps_k}+\\
&+\eta\sum_{k<j}^{\l}\frac{K^+_{1a}(\eps_j)\ell_{aj}\ell_{ak}K^-_{1a}(\eps_j)}{\eps_j-\eps_k}+\eta K^+_{2a}(\eps_j)\ell_{aj}K^-_{1a}(\eps_j)+\\
&+\eta\sum_{k=1}^{\l}\frac{K^+_{1a}(\eps_j)\ell_{aj}K^-_{1a}(\eps_j)\ell_{ak}}{\eps_j+\eps_k}+\eta K^+_{1a}(\eps_j)\ell_{aj}K^-_{2a}(\eps_j)\Bigg]+o(\eta^2).
\end{aligned}
\end{equation*}
One can check that
\begin{equation*}
{\rm tr}_a\left(K^+_{1a}(\eps_j)\ell_{aj}K^-_{1a}(\eps_j)\right)=0,
\end{equation*}
and
\begin{equation*}
{\rm tr}_a\left(K^+_{1a}(\eps_j)\ell_{ak}\ell_{aj}K^-_{1a}(\eps_j)\right)={\rm tr}_a\left(K^+_{1a}(\eps_j)\ell_{aj}\ell_{ak}K^-_{1a}(\eps_j)\right).
\end{equation*}
Thus, we have
\begin{equation*}
\begin{aligned}
\tau_j=&\sum_{k\neq j}^{\l}\frac{{\rm tr}_a\left(K^+_{1a}(\eps_j)\ell_{ak}\ell_{aj}K^-_{1a}(\eps_j)\right)}{\eps_j-\eps_k}+
\sum_{k=1}^{\l}\frac{{\rm tr}_a\left(K^+_{1a}(\eps_j)\ell_{aj}K^-_{1a}(\eps_j)\ell_{ak}\right)}{\eps_j+\eps_k}+\\
&+{\rm tr}_a\left( K^+_{2a}(\eps_j)\ell_{aj}K^-_{1a}(\eps_j)\right)+{\rm tr}_a\left(K^+_{1a}(\eps_j)\ell_{aj}K^-_{2a}(\eps_j)\right).
\end{aligned}
\end{equation*}
Compute the traces:
\begin{equation*}
\begin{aligned}
{\rm tr}_a\left(K^+_{1a}(\eps_j)\ell_{ak}\ell_{aj}K^-_{1a}(\eps_j)\right)=&\left((1+\psi\phi)\eps_j^2-\xi^2\right)\left(2S_j^z S_k^z+S_j^+S_k^-+S_j^-S_k^+\right),\\
{\rm tr}_a\left(K^+_{1a}(\eps_j)\ell_{aj}K^-_{1a}(\eps_j)\ell_{ak}\right)=&\,2(\eps_j+\xi)(\eps_j-\xi)S_j^zS_k^z-(\eps_j-\xi)^2S_j^+S_k^--(\eps_j+\xi)^2S_j^-S_k^++\\
&+2\psi\eps_j\left((\eps_j+\xi)S_j^zS_k^++(\eps_j-\xi)S_j^+S_k^z\right)+2\phi\eps_j\left((\eps_j+\xi)S_j^-S_k^z+(\eps_j-\xi)S_j^zS_k^-\right)+\\
&+\eps_j^2\left(\psi^2 S_j^+ S_k^+ + \phi^2 S_j^- S_k^--2\psi\phi S_j^z S_k^z\right),\\
{\rm tr}_a\left( K^+_{2a}(\eps_j)\ell_{aj}K^-_{1a}(\eps_j)\right)=&\left(2\alpha\eps_j-\xi+(\lambda\psi-\gamma\phi)\eps_j^2\right)S_j^z
+\left((\alpha\psi-\gamma\xi)\eps_j-\frac{\psi}{2}\xi+\gamma\eps_j^2\right)S_j^++\\
&+\left((\alpha\phi-\lambda\xi)\eps_j-\frac{\phi}{2}\xi-\lambda\eps_j^2\right)S_j^-,\\
{\rm tr}_a\left(K^+_{1a}(\eps_j)\ell_{aj}K^-_{2a}(\eps_j)\right)=&\left(2\beta\eps_j-\xi+(\phi\delta-\psi\mu)\eps_j^2\right)S_j^z
+\left((\beta\psi+\delta\xi)\eps_j-\frac{\psi}{2}\xi-\delta\eps_j^2\right)S_j^++\\
&+\left((\beta\phi+\mu\xi)\eps_j-\frac{\phi}{2}\xi+\mu\eps_j^2\right)S_j^-.
\end{aligned}
\end{equation*}
The sum of these four terms leads to a family of conserved operators for the open, rational Richardson-Gaudin model:
\begin{equation}\label{tau}
\begin{aligned}
\tau_j=&\sum_{k\neq j}^{\l}\frac{(1+\psi\phi)\eps_j^2-\xi^2}{\eps_j-\eps_k}\left[2S_j^z S_k^z+S_j^+S_k^-+S_j^-S_k^+\right]+\\
&+\sum_{k=1}^{\l}\frac{1}{\eps_j+\eps_k}\bigg[2(\eps_j+\xi)(\eps_j-\xi)S_j^zS_k^z-(\eps_j-\xi)^2S_j^+S_k^--(\eps_j+\xi)^2S_j^-S_k^++\\
&+2\psi\eps_j\left((\eps_j+\xi)S_j^zS_k^++(\eps_j-\xi)S_j^+S_k^z\right)+2\phi\eps_j\left((\eps_j+\xi)S_j^-S_k^z+(\eps_j-\xi)S_j^zS_k^-\right)+\\
&+\eps_j^2\left(\psi^2 S_j^+ S_k^+ + \phi^2 S_j^- S_k^--2\psi\phi S_j^z S_k^z\right)\bigg]+\\
&+\left[2(\alpha+\beta)\eps_j-2\xi+\psi(\lambda-\mu)\eps_j^2-\phi(\gamma-\delta)\eps_j^2\right]S_j^z+\\
&+\left[\psi(\alpha+\beta)\eps_j-\xi(\gamma-\delta)\eps_j-\psi\xi+(\gamma-\delta)\eps_j^2\right]S_j^++\\
&+\left[\phi(\alpha+\beta)\eps_j-\xi(\lambda-\mu)\eps_j-\phi\xi-(\lambda-\mu)\eps_j^2\right]S_j^-.
\end{aligned}
\end{equation}


\subsection{The second family of conserved operators}

Note that we have only considered one of two possible families of the conserved operators. The second family is constructed as follows from the transfer matrix (\ref{t-b}):
\begin{equation*}
\lim_{u\rightarrow-\eps_j}(u+\eps_j)t(u)=\eta^2\tilde{\tau}_j+o(\eta^2).
\end{equation*}
Here we show that these are equivalent conserved operators, i.e., $\tilde{\tau}_j=-\tau_j$.
Let
\begin{equation*}
t(u,\vec{\eps})={\rm tr}_a\Big(K^+_a(u)L_{a\l}(u-\eps_{\l}) \ldots
L_{a1}(u-\eps_1)K_a^-(u)L_{a1}(u+\eps_1)\ldots L_{a\l}(u+\eps_{\l})\Big).
\end{equation*}
Consider $t(u,\vec{\eps})^T$, where $T=t_1\ldots t_{\l}$ denotes a transpose over all spaces. Using 
\begin{equation*}
\left({\rm tr}_a A_a\right)^{t_1\ldots  t_{\l}}={\rm tr}_a\left(A_a^{t_1\ldots t_{\l}}\right)={\rm
tr}_a\left(A_a^{t_at_1\ldots t_{\l}}\right),
\end{equation*}
the fact that Lax operators are symmetric
\begin{equation*}
L_{aj}(u)^T=\frac{1}{u}\begin{pmatrix}
                                         u+\eta (S_j^z)^T & \eta (S_j^+)^T\\
                                         \eta (S_j^-)^T & u-\eta (S_j^z)^T
                                         \end{pmatrix}=\frac{1}{u}\begin{pmatrix}
                                                                                     u+\eta S_j^z & \eta S_j^-\\
                                                                                     \eta S_j^+ & u-\eta S_j^z
                                                                                     \end{pmatrix}=L_{aj}(u),
\end{equation*}
and an observation that $K^+(u)^T=K^+(u)|_{\psi^+\leftrightarrow\phi^+}$ and $K^-(u)^T=K^-(u)|_{\psi^-\leftrightarrow\phi^-}$, we obtain
\begin{equation*}
\begin{aligned}
t(u,\vec{\eps})^T=&\,{\rm tr}_a\Big(L_{a\l}(u+\eps_{\l})\ldots L_{a1}(u+\eps_1)K_a^-(u)^T
L_{a1}(u-\eps_1)\ldots L_{a\l}(u-\eps_N)K^+_a(u)^T\Big)=\\
=&\,{\rm tr}_a\Big(K^+_a(u)L_{a\l}(u+\eps_{\l})\ldots L_{a1}(u+\eps_1)K_a^-(u)L_{a1}(u-\eps_1)\ldots L_{a\l}(u-\eps_{\l})\Big)|_{\psi^+\leftrightarrow\phi^+,\psi^-\leftrightarrow\phi^-}=\\
=&\,t(u,-\vec{\eps})|_{\psi^+\leftrightarrow\phi^+,\psi^-\leftrightarrow\phi^-}.
\end{aligned}
\end{equation*}
Thus, we obtain the following equality:
\begin{equation*}
t(u,\vec{\eps})=t(u,-\vec{\eps})^T|_{\psi^+\leftrightarrow\phi^+,\psi^-\leftrightarrow\phi^-}.
\end{equation*}
It follows that
\begin{equation*}
\lim_{u\rightarrow-\eps_j}(u+\eps_j)t(u,\vec{\eps})=\lim_{u\rightarrow-\eps_j}(u-(-\eps_j))t(u,-\vec{\eps})^T|_{\psi^+\leftrightarrow\phi^+,\psi^-\leftrightarrow\phi^-}.
\end{equation*}
Thus, we have
\begin{equation*}
\tilde{\tau}_j(\vec{\eps})=\tau_j(-\vec{\eps})^T|_{\psi^+\leftrightarrow\phi^+,\psi^-\leftrightarrow\phi^-}=
\tau_j(-\vec{\eps})^T|_{\psi\leftrightarrow\phi,\gamma\leftrightarrow\lambda,\delta\leftrightarrow\mu}. 
\end{equation*}
Computing  $\tilde{\tau}_j(\vec{\eps})=\tau_j(-\vec{\eps})^T|_{\psi\leftrightarrow\phi,\gamma\leftrightarrow\lambda,\delta\leftrightarrow\mu}$ from (\ref{tau}) we obtain
that  $\tilde{\tau}_j(\vec{\eps})=-\tau_j(\vec{\eps})$.


\subsection{The case when one $K$-matrix is diagonal}

Hereafter, we will only consider the spin-$1/2$ representation of this algebra acting on $V$:
\begin{equation*}
S^+=\begin{pmatrix}
              0 & 1 \\
              0 & 0
             \end{pmatrix}, \
S^-=\begin{pmatrix}
              0 & 0 \\
              1 & 0
             \end{pmatrix},\ 
S^z=\dfrac{1}{2}\begin{pmatrix}
                1 & 0 \\
                0 & -1
                \end{pmatrix}.
\end{equation*}
\begin{remark}
Note that in this instance $L_{aj}(u)=u^{-1}R_{aj}(u-\eta/2)$.
\end{remark}
It now turns out that six of the parameters appearing in (\ref{tau}) are superfluous and can be eliminated by appropriate basis transformations and redefinitions of variables. First note that we can set $\beta=0$ without loss of generality, since the dependence of (\ref{tau}) on $\alpha$ and $\beta$ is only through the sum $\alpha+\beta$. Next, the Lax operator is invariant under the local basis transformations, i.e. 
\begin{equation*}
X_a X_j L_{aj}(u) X_a^{-1} X_j^{-1}=L_{aj}(u)
\end{equation*}
for any invertible $X\in\text{End}(\c^2)$. 
Thus we can almost always choose a basis in which one of the $K$-matrices is diagonal. (The case when a $K$-matrix is not diagonalisable has been discussed in \cite{cms14}). For our purposes, we assume that $K^-(u)$ is
diagonal, so that
\begin{align*}
K^-(u)=&\begin{pmatrix}
                 \xi^- +u-\eta/2 & 0 \\
                 0 & \xi^- -u+\eta/2
                 \end{pmatrix}, \\
K^+(u)=&\begin{pmatrix}
                 \xi^++u+\eta/2 & \psi^+(u+\eta/2)\\
                 \phi^+(u+\eta/2) & \xi^+-u-\eta/2
                 \end{pmatrix}, 
\end{align*}
For the expansion (\ref{eta}) this means that $\psi=\phi= \delta=\mu=0$ . Substituting these into (\ref{tau}) we obtain
\begin{equation*}
\begin{aligned}
\tau_j=&\,(\eps_j-\xi)(\eps_j+\xi)\Bigg[\sum_{k\neq j}^{\l}\left(\frac{1}{\eps_j-\eps_k}+\frac{1}{\eps_j+\eps_k}\right)2S_j^z S_k^z+
\sum_{k\neq j}^{\l}\left(\frac{1}{\eps_j-\eps_k}-\frac{1}{\eps_j+\eps_k}\frac{\eps_j-\xi}{\eps_j+\xi}\right)S_j^+ S_k^-+\\
&+\sum_{k\neq j}^{\l}\left(\frac{1}{\eps_j-\eps_k}-\frac{1}{\eps_j+\eps_k}\frac{\eps_j+\xi}{\eps_j-\xi}\right)S_j^- S_k^++\frac{1}{2\eps_j}2(S_j^z)^2-
\frac{1}{2\eps_j}\frac{\eps_j-\xi}{\eps_j+\xi}S_j^+S_j^--\frac{1}{2\eps_j}\frac{\eps_j+\xi}{\eps_j-\xi}S_j^-S_j^++\\
&+\frac{2\alpha\eps_j}{\eps_j^2-\xi^2}S_j^z-\frac{2\xi}{\eps_j^2-\xi^2}S_j^z+\frac{\gamma\eps_j}{\eps_j+\xi}S_j^+-\frac{\lambda\eps_j}{\eps_j-\xi}S_j^-\Bigg].
\end{aligned}
\end{equation*}

Finally we may set $\xi=0$ without loss of generality, although this is more technical to establish.
Using the properties of the spin-$1/2$ representation, namely
\begin{equation*}
S^+S^-=\dfrac{1}{2}I+S^z,\ S^-S^+=\dfrac{1}{2}I-S^z,\ (S^z)^2=\dfrac{1}{4}I,
\end{equation*}
and the identities
\begin{equation*}
\begin{aligned}
\frac{1}{\eps_j-\eps_k}-\frac{1}{\eps_j+\eps_k}\frac{\eps_j-\xi}{\eps_j+\xi}=&\frac{2\eps_j(\eps_k+\xi)}{(\eps_j^2-\eps_k^2)(\eps_j+\xi)},\\
\frac{1}{\eps_j-\eps_k}-\frac{1}{\eps_j+\eps_k}\frac{\eps_j+\xi}{\eps_j-\xi}=&\frac{2\eps_j(\eps_k-\xi)}{(\eps_j^2-\eps_k^2)(\eps_j-\xi)}
\end{aligned}
\end{equation*}
we may simplify the expression for $\tau_j$ to write
\begin{equation}\label{tau-semi-diag*}
\begin{aligned}
\frac{\eps_j\tau_j}{(\eps_j-\xi)(\eps_j+\xi)}=&\sum_{k\neq j}^{\l}\frac{4\eps_j^2}{\eps_j^2-\eps_k^2}S_j^z S_k^z+
\sum_{k\neq j}^{\l}\frac{2\eps_j^2}{\eps_j^2-\eps_k^2}\left(\frac{\eps_k+\xi}{\eps_j+\xi}S_j^+S_k^-+\frac{\eps_k-\xi}{\eps_j-\xi}S_j^-S_k^+\right)+\\
&+\frac{2\alpha\eps_j^2}{\eps_j^2-\xi^2}S_j^z+\frac{\gamma\eps_j^2}{\eps_j+\xi}S_j^+-\frac{\lambda\eps_j^2}{\eps_j-\xi}S_j^-
+\frac{1}{4}I-\frac{1}{2}\frac{\eps_j^2+\xi^2}{\eps_j^2-\xi^2}I.
\end{aligned}
\end{equation}
Consider the following local transformation on the $j$th space in the tensor product:
\begin{equation*}
 U_j=\text{diag}\left(\sqrt{\frac{\eps_j+\xi}{\eps_j-\xi}},1\right).
\end{equation*}
Under these transformations we have
\begin{equation*}
\begin{aligned}
U_j S_j^z U_j^{-1}=&\ S_j^z,\\
U_j S_j^+ U_j^{-1}=&\ \sqrt{\frac{\eps_j+\xi}{\eps_j-\xi}} S_j^+,\\
U_j S_j^- U_j^{-1}=&\ \sqrt{\frac{\eps_j-\xi}{\eps_j+\xi}}S_j^-. 
\end{aligned}
\end{equation*}
Define
\begin{equation*}
\begin{aligned}
\tau_j^{(1)}=&\sum_{k\neq j}^{\l}\frac{4\eps_j^2}{\eps_j^2-\eps_k^2}S_j^z S_k^z+
\sum_{k\neq j}^{\l}\frac{2\eps_j^2}{\eps_j^2-\eps_k^2}\left(\frac{\eps_k+\xi}{\eps_j+\xi}S_j^+S_k^-+\frac{\eps_k-\xi}{\eps_j-\xi}S_j^-S_k^+\right)
+\frac{2\alpha\eps_j^2}{\eps_j^2-\xi^2}S_j^z+\\
&+\frac{\gamma\eps_j^2}{\eps_j+\xi} S_j^+-\frac{\lambda\eps_j^2}{\eps_j-\xi} S_j^-.
\end{aligned}
\end{equation*}
We see that, up to a constant term, it is the same expression as (\ref{tau-semi-diag*}).
Under the global transformation $U=U_1 U_2\ldots U_\l$ we define
\begin{equation*}
\begin{aligned}
\tau_j^{(2)}&=U \tau_j^{(1)} U^{-1}= \\
&=\sum_{k\neq j}^{\l}\frac{4\eps_j^2}{\eps_j^2-\eps_k^2}S_j^z S_k^z+
\sum_{k\neq j}^{\l}\frac{2\eps_j^2}{\eps_j^2-\eps_k^2}\frac{\sqrt{\eps_k^2-\xi^2}}{\sqrt{\eps_j^2-\xi^2}}(S_j^+S_k^-+S_j^-S_k^+)
+\frac{2\alpha\eps_j^2}{\eps_j^2-\xi^2}S_j^z+\\
&+\frac{\gamma\eps_j^2}{\sqrt{\eps_j^2-\xi^2}} S_j^+-\frac{\lambda\eps_j^2}{\sqrt{\eps_j^2-\xi^2}} S_j^-.
\end{aligned}
\end{equation*}
Next simply rescale to obtain
\begin{equation*}
\begin{aligned}
\tau_j^{(3)}=& \dfrac{\eps_j^2-\xi^2}{\eps_j^2} \tau^{(2)}_j= \\
 =&\sum_{k\neq j}^{\l}\frac{4(\eps_j^2-\xi^2)}{\eps_j^2-\eps_k^2}S_j^z S_k^z+
\sum_{k\neq j}^{\l}\frac{2\sqrt{\eps_j^2-\xi^2}\sqrt{\eps_k^2-\xi^2}}{\eps_j^2-\eps_k^2}(S_j^+S_k^-+S_j^-S_k^+)+2\alpha S_j^z+\\
&+\gamma\sqrt{\eps_j^2-\xi^2} S_j^+-\lambda\sqrt{\eps_j^2-\xi^2} S_j^-.
\end{aligned}
\end{equation*}
Now we apply a change of variables  $\eps_j\mapsto\sqrt{\eps_j^2+\xi^2}$ to obtain 
\begin{equation}\label{tau-final}
\begin{aligned}
\tau_j^{*}=&\sum_{k\neq j}^{\l}\frac{4\eps_j^2}{\eps_j^2-\eps_k^2}S_j^z S_k^z+
\sum_{k\neq j}^{\l}\frac{2\eps_j\eps_k}{\eps_j^2-\eps_k^2}(S_j^+S_k^-+S_j^-S_k^+)+2\alpha S_j^z+\\
&+\gamma\eps_j S_j^+-\lambda\eps_j S_j^- =  \\
=& \left.\frac{\eps_j\tau_j}{(\eps_j-\xi)(\eps_j+\xi)}\right|_{\xi=0}+\dfrac{1}{4}I.
\end{aligned}
\end{equation}
This affirms that we may also set $\xi=0$ without loss of generality.

We refer to the set of mutually commuting conserved operators $\{\tau_j^*:j=1,...,\l\}$ as the open, rational Richardson-Gaudin system in the spin-1/2 case. Note that the coefficients of the $S_j^zS_k^z$ terms in (\ref{tau-final}) are not antisymmetric with respect to the interchange of indices $j$ and $k$. This distinguishes this set of commuting operators from those obtained by the Gaudin algebra approach \cite{rdo10,vdv14,cdvv15}



\subsection{Hamiltonian}

Let us now construct the Hamiltonian (\ref{H}) from these conserved operators. 
Consider
\begin{equation*}
\begin{aligned}
\sum_{j=1}^{\l}\eps_j^{-2}\tau_j^*=&-2\sum_{j,k:
j<k}\eps_j^{-1}\eps_k^{-1}(S_j^+S_k^-+S_j^-S_k^+)+2\alpha\sum_{j=1}^{\l}\eps_j^{-2}
S_j^z+\\
&+\gamma\sum_{j=1}^{\l}\eps_j^{-1} S_j^+-\lambda\sum_{j=1}^{\l}\eps_j^{-1}
S_j^-.
\end{aligned}
\end{equation*}
Setting $\lambda=-\gamma$, and making the
change of variable $z_j=\eps_j^{-1}$ we obtain 
\begin{align}
H&=\frac{1}{2\alpha}\sum_{j=1}^{\l}\eps_j^{-2}\tau_j^* = \nonumber \\
&=\sum_{j=1}^{\l}z_j^2 S_j^z-\frac{1}{\alpha}\sum_{j,k: j<k}z_j z_k (S_j^+S_k^-+S_j^-S_k^+)
+\frac{\gamma}{2\alpha}\sum_{j=1}^{\l}z_j(S_j^+ +S_j^-). \label{hamiltonian}
\end{align}
We see that (\ref{hamiltonian}) is equivalent to (\ref{H}) by identifying $\alpha=G^{-1}$ and ${\gamma}=2\Gamma G^{-1}$.



\section{Eigenvalues, Bethe Ansatz Equations and the energy spectrum}

We now turn to investigating the eigenvalues of the conserved operators, making use of the results
of Wang et al. \cite{hcyy15}).

\subsection{Eigenvalues}

Rewrite the $K$-matrices (\ref{K-}) and (\ref{K+}) in the following form (using the notation from
\cite{hcyy15}):
\begin{equation*}
\begin{aligned}
K^-(u)=&\,\xi^-+(u-\eta/2)\vec{h}_1\cdot\vec{\sigma}\ \ \text{with}\ \ \vec{h}_1=\left(\frac{\psi^-+\phi^-}{2},\frac{i(\psi^--\phi^-)}{2},1\right),\\
K^+(u)=&\,\xi^++(u+\eta/2)\vec{h}_2\cdot\vec{\sigma}\ \ \text{with}\ \ \vec{h}_2=\left(\frac{\psi^++\phi^+}{2},\frac{i(\psi^+-\phi^+)}{2},1\right),
\end{aligned}
\end{equation*}
where $\vec{\sigma}=(\sigma^x,\sigma^y,\sigma^z)$ are the Pauli matrices. To match the notation from \cite{wycs15} we need to normalise the vectors $\vec{h}_1$ and $\vec{h}_2$
\begin{equation*}
\vec{h}_1^0=\frac{\vec{h}_1}{\sqrt{\psi^-\phi^-+1}},\ \ \vec{h}_2^0=\frac{\vec{h}_2}{\sqrt{\psi^+\phi^++1}}.
\end{equation*}
The $K$-matrices can then be written as
\begin{equation*}
\begin{aligned}
K^-(u)=&\sqrt{\psi^-\phi^-+1}\left(\frac{\xi^-}{\sqrt{\psi^-\phi^-+1}}+(u-\eta/2)\vec{h}_1^0\cdot\vec{\sigma}\right),\\
K^+(u)=&\sqrt{\psi^+\phi^++1}\left(\frac{\xi^+}{\sqrt{\psi^+\phi^++1}}+(u+\eta/2)\vec{h}_2^0\cdot\vec{\sigma}\right).
\end{aligned}
\end{equation*}
Let $\{v_k\ |\  k=1,2,\ldots \l\}$ denote a set of parameters that will be utilised to determine the
eigenvalues of the transfer matrix (\ref{t-b}). From \cite{hcyy15}, the formula for the eigenvalues of (\ref{t-b}) is
\begin{equation*}
\Lambda(u)=\sqrt{\psi^-\phi^-+1}\sqrt{\psi^+\phi^++1}\Bigg[a(u)\frac{Q(u+\eta)}{Q(u)}+d(u)\frac{Q(u-\eta)}{Q(u)}+c(u-\eta/2)(u+\eta/2)\frac{F(u)}{Q(u)}\Bigg],
\end{equation*}
where
\begin{equation*}
\begin{aligned}
&Q(u)=\prod_{i=1}^{\l}(u-v_i)(u+v_i),\\
&a(u)=\frac{2u-\eta}{2u}\left(u+\frac{\xi^-}{\sqrt{\psi^-\phi^-+1}}+\eta/2\right)\left(u+\frac{\xi^+}{\sqrt{\psi^+\phi^++1}}+\eta/2\right)
\prod_{l=1}^{\l}\frac{(u-\eps_l-\eta/2)(u+\eps_l-\eta/2)}{(u-\eps_l)(u+\eps_l)},\\
&d(u)=\frac{2u+\eta}{2u}\left(u-\frac{\xi^-}{\sqrt{\psi^-\phi^-+1}}-\eta/2\right)\left(u-\frac{\xi^+}{\sqrt{\psi^+\phi^++1}}-\eta/2\right)
\prod_{l=1}^{\l}\frac{(u-\eps_l+\eta/2)(u+\eps_l+\eta/2)}{(u-\eps_l)(u+\eps_l)},\\
&F(u)=\prod_{i=1}^{\l}\frac{(u+\eps_i-\eta/2)(u-\eps_i-\eta/2)(u+\eps_i+\eta/2)(u-\eps_i+\eta/2)}{(u-\eps_i)(u+\eps_i)},\\
&c=2\left(\vec{h}^0_1\cdot\vec{h}^0_2-1\right).
\end{aligned}
\end{equation*}
The constant $c$ can be computed as
\begin{equation*}
c=2\left(\frac{\vec{h}_1\cdot\vec{h}_2}{\sqrt{\psi^-\phi^-+1}\sqrt{\psi^+\phi^++1}}-1\right)
=2\left(\frac{\frac{1}{2}(\psi^-\phi^++\phi^-\psi^+)+1}{\sqrt{\psi^-\phi^-+1}\sqrt{\psi^+\phi^++1}}-1\right).
\end{equation*}
Finally, we obtain
\begin{equation}\label{b-ev}
\begin{aligned}
\Lambda(u)=&\sqrt{\psi^-\phi^-+1}\sqrt{\psi^+\phi^++1}\Bigg[a(u)\prod_{i=1}^{\l}\frac{(u-v_i+\eta)(u+v_i+\eta)}{(u-v_i)(u+v_i)}
+d(u)\prod_{i=1}^{\l}\frac{(u-v_i-\eta)(u+v_i-\eta)}{(u-v_i)(u+v_i)}+\\
&+c(u^2-\eta^2/4)\prod_{i=1}^{\l}\frac{((u+\eps_i)^2-\eta^2/4)((u-\eps_i)^2-\eta^2/4)}{(u^2-\eps_i^2)(u^2-v_i^2)}\Bigg],
\end{aligned}
\end{equation}
where
\begin{equation*}
\begin{aligned}
&a(u)=\frac{2u-\eta}{2u}\left(u+\frac{\xi^-}{\sqrt{\psi^-\phi^-+1}}+\frac{\eta}{2}\right)\left(u+\frac{\xi^+}{\sqrt{\psi^+\phi^++1}}+\frac{\eta}{2}\right)
\prod_{l=1}^{\l}\frac{(u-\eps_l-\eta/2)(u+\eps_l-\eta/2)}{(u-\eps_l)(u+\eps_l)},\\
&d(u)=\frac{2u+\eta}{2u}\left(u-\frac{\xi^-}{\sqrt{\psi^-\phi^-+1}}-\frac{\eta}{2}\right)\left(u-\frac{\xi^+}{\sqrt{\psi^+\phi^++1}}-\frac{\eta}{2}\right)
\prod_{l=1}^{\l}\frac{(u-\eps_l+\eta/2)(u+\eps_l+\eta/2)}{(u-\eps_l)(u+\eps_l)}.
\end{aligned}
\end{equation*}

\subsection{Quasi-classical limit of the eigenvalues}

The eigenvalues in the quasi-classical limit ($\eta\rightarrow 0$) are constructed as follows:
\begin{equation*}
\lim_{u\rightarrow\eps_j}(u-\eps_j)\Lambda(u)=\eta^2{\lambda}_j+o(\eta^2).
\end{equation*}
We compute this limit assuming the same dependencies (\ref{eta}) as for the conserved operators:
\begin{equation*}
\begin{aligned}
\lim_{u\rightarrow\eps_j}(u-\eps_j)\prod_{l=1}^{\l}\frac{(u+\eps_l-\eta/2)(u-\eps_l-\eta/2)}{(u-\eps_l)(u+\eps_l)}= &
-\frac{\eta}{2}+\frac{\eta^2}{4}\left[\frac{1}{2\eps_j}+\sum_{k\neq j}^{\l}\left(\frac{1}{\eps_j-\eps_k}+\frac{1}{\eps_j+\eps_k}\right)\right]+o(\eta^2),\\
\lim_{u\rightarrow\eps_j}(u-\eps_j)\prod_{l=1}^{\l}\frac{(u+\eps_l+\eta/2)(u-\eps_l+\eta/2)}{(u-\eps_l)(u+\eps_l)}= &
\,\frac{\eta}{2}+\frac{\eta^2}{4}\left[\frac{1}{2\eps_j}+\sum_{k\neq j}^{\l}\left(\frac{1}{\eps_j-\eps_k}+\frac{1}{\eps_j+\eps_k}\right)\right]+o(\eta^2),
\end{aligned}
\end{equation*}
\begin{equation*}
\begin{aligned}
\sqrt{\psi^-\phi^-+1}= &\sqrt{\psi\phi+1}\left(1+\frac{\eta}{2}\frac{\mu\psi+\delta\phi}{\psi\phi+1}\right)+o(\eta),\\
\sqrt{\psi^+\phi^++1}= &\sqrt{\psi\phi+1}\left(1+\frac{\eta}{2}\frac{\lambda\psi+\gamma\phi}{\psi\phi+1}\right)+o(\eta).
\end{aligned}
\end{equation*}
Consider the expansion of the first two terms in (\ref{b-ev}) up to the second order in $\eta$:
\begin{equation*}
\begin{aligned}
\lim_{u\rightarrow\eps_j}(u-\eps_j)&\sqrt{\psi^-\phi^-+1}\sqrt{\psi^+\phi^++1}\,a(u)= -\frac{\eta}{2}(\eps_j^2(\psi\phi+1)-\xi^2)
-\frac{\eta^2}{2}\Bigg(\frac{\xi^2}{2\eps_j}+\frac{\eps_j}{2}(\psi\phi+1)+\\
&+\frac{\eps_j^2}{2}((\lambda+\mu)\psi+(\gamma+\delta)\phi)+(\alpha+\beta)\eps_j\sqrt{\psi\phi+1}
-\frac{\xi\eps_j}{2\sqrt{\psi\phi+1}}((\lambda-\mu)\psi+(\gamma-\delta)\phi)-\xi(\alpha-\beta)\Bigg)+\\
&+\frac{\eta^2}{4}(\eps_j^2(\psi\phi+1)-\xi^2)\left(\frac{1}{2\eps_j}+\sum_{k\neq j}^{\l}\left(\frac{1}{\eps_j-\eps_k}+\frac{1}{\eps_j+\eps_k}\right)\right)+o(\eta^2),
\end{aligned}
\end{equation*}
\begin{equation*}
\begin{aligned}
\lim_{u\rightarrow\eps_j}(u-\eps_j)&\sqrt{\psi^-\phi^-+1}\sqrt{\psi^+\phi^++1}\,d(u)= \frac{\eta}{2}(\eps_j^2(\psi\phi+1)-\xi^2)
+\frac{\eta^2}{2}\Bigg(-\frac{\xi^2}{2\eps_j}-\frac{\eps_j}{2}(\psi\phi+1)+\\
&+\frac{\eps_j^2}{2}((\lambda+\mu)\psi+(\gamma+\delta)\phi)-(\alpha+\beta)\eps_j\sqrt{\psi\phi+1}
+\frac{\xi\eps_j}{2\sqrt{\psi\phi+1}}((\lambda-\mu)\psi+(\gamma-\delta)\phi)-\xi(\alpha-\beta)\Bigg)+\\
&+\frac{\eta^2}{4}(\eps_j^2(\psi\phi+1)-\xi^2)\left(\frac{1}{2\eps_j}+\sum_{k\neq j}^{\l}\left(\frac{1}{\eps_j-\eps_k}+\frac{1}{\eps_j+\eps_k}\right)\right)+o(\eta^2).
\end{aligned}
\end{equation*}
Also
\begin{equation*}
\begin{aligned}
\prod_{i=1}^{\l}\frac{(u-v_i+\eta)(u+v_i+\eta)}{(u-v_i)(u+v_i)}= &\,1+\eta\sum_{i=1}^{\l}\left(\frac{1}{u-v_i}+\frac{1}{u+v_i}\right)+o(\eta),\\
\prod_{i=1}^{\l}\frac{(u-v_i-\eta)(u+v_i-\eta)}{(u-v_i)(u+v_i)}= &\,1-\eta\sum_{i=1}^{\l}\left(\frac{1}{u-v_i}+\frac{1}{u+v_i}\right)+o(\eta).
\end{aligned}
\end{equation*}
Combining these calculations then leads to
\begin{equation*}
\begin{aligned}
\lim_{u\rightarrow\eps_j}(u-\eps_j)&\sqrt{\psi^-\phi^-+1}\sqrt{\psi^+\phi^++1}\,a(u)\prod_{i=1}^{\l}\frac{(u-v_i+\eta)(u+v_i+\eta)}{(u-v_i)(u+v_i)}=
-\frac{\eta}{2}(\eps_j^2(\psi\phi+1)-\xi^2)+\\
&+\frac{\eta^2}{2}\Bigg[-(\eps_j^2(\psi\phi+1)-\xi^2)\sum_{i=1}^{\l}\frac{2\eps_j}{\eps_j^2-v_i^2}
+\frac{1}{2}(\eps_j^2(\psi\phi+1)-\xi^2)\sum_{k\neq j}^{\l}\frac{2\eps_j}{\eps^2_j-\eps^2_k}-\\
&-\frac{\eps_j}{4}(\psi\phi+1)-\frac{3\xi^2}{4\eps_j}-\frac{\eps_j^2}{2}((\lambda+\mu)\psi+(\gamma+\delta)\phi)-(\alpha+\beta)\eps_j\sqrt{\psi\phi+1}+\\
&+\frac{\xi\eps_j}{2\sqrt{\psi\phi+1}}((\lambda-\mu)\psi+(\gamma-\delta)\phi)+\xi(\alpha-\beta)\Bigg]+o(\eta^2),
\end{aligned}
\end{equation*}
\begin{equation*}
\begin{aligned}
\lim_{u\rightarrow\eps_j}(u-\eps_j)&\sqrt{\psi^-\phi^-+1}\sqrt{\psi^+\phi^++1}\,d(u)\prod_{i=1}^{\l}\frac{(u-v_i-\eta)(u+v_i-\eta)}{(u-v_i)(u+v_i)}=
\frac{\eta}{2}(\eps_j^2(\psi\phi+1)-\xi^2)+\\
&+\frac{\eta^2}{2}\Bigg[-(\eps_j^2(\psi\phi+1)-\xi^2)\sum_{i=1}^{\l}\frac{2\eps_j}{\eps_j^2-v_i^2}
+\frac{1}{2}(\eps_j^2(\psi\phi+1)-\xi^2)\sum_{k\neq j}^{\l}\frac{2\eps_j}{\eps^2_j-\eps^2_k}-\\
&-\frac{\eps_j}{4}(\psi\phi+1)-\frac{3\xi^2}{4\eps_j}+\frac{\eps_j^2}{2}((\lambda+\mu)\psi+(\gamma+\delta)\phi)-(\alpha+\beta)\eps_j\sqrt{\psi\phi+1}+\\
&+\frac{\xi\eps_j}{2\sqrt{\psi\phi+1}}((\lambda-\mu)\psi+(\gamma-\delta)\phi)-\xi(\alpha-\beta)\Bigg]+o(\eta^2).
\end{aligned}
\end{equation*}
Finally, the sum of the first two terms in (\ref{b-ev}) can be expressed as
\begin{equation*}
\begin{aligned}
&\lim_{u\rightarrow\eps_j}(u-\eps_j)\sqrt{\psi^-\phi^-+1}\sqrt{\psi^+\phi^++1}\Bigg[a(u)\prod_{i=1}^{\l}\frac{(u-v_i+\eta)(u+v_i+\eta)}{(u-v_i)(u+v_i)}
+d(u)\prod_{i=1}^{\l}\frac{(u-v_i-\eta)(u+v_i-\eta)}{(u-v_i)(u+v_i)}\Bigg]=\\
=&\,\eta^2\Bigg[\frac{(\eps_j^2(\psi\phi+1)-\xi^2)}{\eps_j}
\left(\sum_{k\neq j}^{\l}\frac{\eps_j^2}{\eps_j^2-\eps_k^2}-\sum_{i=1}^{\l}\frac{2\eps_j^2}{\eps_j^2-v_i^2}+\frac{3}{4}\right)
-\eps_j(\psi\phi+1)-(\alpha+\beta)\eps_j\sqrt{\psi\phi+1}+\\
&+\frac{\xi\eps_j}{2\sqrt{\psi\phi+1}}((\lambda-\mu)\psi+(\gamma-\delta)\phi)\Bigg]+o(\eta^2).
\end{aligned}
\end{equation*}
The third term of (\ref{b-ev}), reproduced here for convenience,
\begin{equation*}
\lim_{u\rightarrow\eps_j}(u-\eps_j)\left[\sqrt{\psi^-\phi^-+1}\sqrt{\psi^+\phi^++1}\,c(u^2-\eta^2/4)
\prod_{i=1}^{\l}\frac{((u+\eps_i)^2-\eta^2/4)((u-\eps_i)^2-\eta^2/4)}{(u^2-\eps_i^2)(u^2-v_i^2)}\right],
\end{equation*}
is computed as follows.
First, expand the product in powers of $\eta$:
\begin{equation*}
\begin{aligned}
&\prod_{i=1}^{\l}\frac{((u+\eps_i)^2-\eta^2/4)((u-\eps_i)^2-\eta^2/4)}{(u^2-\eps_i^2)(u^2-v_i^2)}=
  \prod_{i=1}^{\l}\frac{(u+\eps_i)^2(u-\eps_i)^2-({\eta^2}/{4})\left((u+\eps_i)^2+(u-\eps_i)^2\right)+o(\eta^2)}{(u+\eps_i)(u+\eps_i)(u^2-v_i^2)}=\\
= &\prod_{i=1}^{\l}\frac{u^2-\eps_i^2}{u^2-v_i^2}
-\frac{\eta^2}{4}\sum_{i,k:i\neq k}^{\l}\frac{u^2-\eps_i^2}{u^2-v_i^2}\frac{1}{u^2-v_k^2}\left(\frac{u+\eps_k}{u-\eps_k}+\frac{u-\eps_k}{u+\eps_k}\right)+o(\eta^2).
\end{aligned}
\end{equation*}
Thus,
\begin{equation*}
\lim_{u\rightarrow\eps_j}(u-\eps_j)\prod_{i=1}^{\l}\frac{((u+\eps_i)^2-\eta^2/4)((u-\eps_i)^2-\eta^2/4)}{(u^2-\eps_i^2)(u^2-v_i^2)}=
-\frac{\eta^2}{4}\sum_{i\neq j}^{\l}\frac{\eps_j^2-\eps_i^2}{\eps_j^2-v_i^2}\frac{2\eps_j}{\eps_j^2-v_j^2}+o(\eta^2).
\end{equation*}
This term already gives the multiple of $\eta^2$, so we just need to consider the constant contribution from the other multiples. Consider
\begin{equation*}
\begin{aligned}
\left.\sqrt{\psi^-\phi^-+1}\sqrt{\psi^+\phi^++1}\,c\right|_{\eta=0}&=(\psi^-\phi^++\phi^-\psi^+)+2-2\sqrt{\psi^-\phi^-+1}\sqrt{\psi^+\phi^++1}|_{\eta=0}=\\
&=2\psi\phi+2-2(\psi\phi+1)=0
\end{aligned}
\end{equation*}
Thus, there will be no contribution in the eigenvalues from the third term in (\ref{b-ev}).

Finally, we obtain the eigenvalues of the conserved operators (\ref{tau}) 
as
\begin{equation}\label{qc-ev}
\begin{aligned}
{\lambda}_j=&\,\frac{(\eps_j^2(\psi\phi+1)-\xi^2)}{\eps_j}
\left(\sum_{k\neq j}^{\l}\frac{\eps_j^2}{\eps_j^2-\eps_k^2}-\sum_{i=1}^{\l}\frac{2\eps_j^2}{\eps_j^2-v_i^2}+\frac{3}{4}\right)
-\eps_j(\psi\phi+1)-(\alpha+\beta)\eps_j\sqrt{\psi\phi+1}+\\
&+\frac{\xi\eps_j}{2\sqrt{\psi\phi+1}}((\lambda-\mu)\psi+(\gamma-\delta)\phi).
\end{aligned}
\end{equation}
\begin{remark}
In view of (\ref{tau-final}), by setting $\beta=\psi=\phi=\delta=\mu=\xi=0$ in (\ref{qc-ev}) we deduce the eigenvalues of $\tau_j^*$ to be 
%
\begin{equation}\label{ev-final}
\lambda_j^*=\sum_{k\neq j}^{\l}\frac{\eps_j^2}{\eps_j^2-\eps_k^2}-\sum_{i=1}^{\l}\frac{2\eps_j^2}{\eps_j^2-v_i^2}-\alpha.
\end{equation}
\end{remark}


\subsection{Bethe Ansatz Equations}

The eigenvalue expression for $\Lambda(u)$ given in (\ref{b-ev}) is undefined for $u=v_k$, for each
$k=1,2,\ldots \l$. Assuming that the $v_k$ are all distinct, analyticity of $\Lambda(u)$ requires that
$\displaystyle{\lim_{u\rightarrow v_k}\Lambda(u)}$ must be finite for each $k=1,2,\ldots,\l.$ This
requirement equates to evaluating the residue of $\Lambda(u)$ at $u=v_k$, and the resulting
constraints on the $v_k$ are referred to as the Bethe Ansatz Equations (BAE).  
The BAE are equivalent to 
\begin{equation}\label{lim}
\lim_{u\rightarrow v_k}(u-v_k)\Lambda(u)=0,
\end{equation}
Compute (\ref{lim}) from (\ref{b-ev}):
\begin{equation}\label{b-bae}
\begin{aligned}
&\frac{2\eta}{v_k}\left(v_k\sqrt{\psi^-\phi^-+1}+\xi^-+\frac{\eta}{2}\sqrt{\psi^-\phi^-+1}\right)
\left(v_k\sqrt{\psi^+\phi^++1}+\xi^++\frac{\eta}{2}\sqrt{\psi^+\phi^++1}\right)\times\\
&\times\prod_{l=1}^{\l}\frac{1}{(v_k-\eps_l+\eta/2)(v_k+\eps_l+\eta/2)}\prod_{i\neq k}^{\l}(v_k-v_i+\eta)(v_k+v_i+\eta)-\\
&-\frac{2\eta}{v_k}\left(v_k\sqrt{\psi^-\phi^-+1}-\xi^--\frac{\eta}{2}\sqrt{\psi^-\phi^-+1}\right)
\left(v_k\sqrt{\psi^+\phi^++1}-\xi^+-\frac{\eta}{2}\sqrt{\psi^+\phi^++1}\right)\times\\
&\times\prod_{l=1}^{\l}\frac{1}{(v_k-\eps_l-\eta/2)(v_k+\eps_l-\eta/2)}\prod_{i\neq k}^{\l}(v_k-v_i-\eta)(v_k+v_i-\eta)+\\
&+\left((\psi^-\phi^++\phi^-\psi^+)+2-2\sqrt{\psi^-\phi^-+1}\sqrt{\psi^+\phi^++1}\right)=0.
\end{aligned}
\end{equation}
\begin{remark}
One can also compute the BAE from
\begin{equation*}
\lim_{u\rightarrow -v_k}(u+v_k)\Lambda(u)=0.
\end{equation*}
This gives the same expression (\ref{b-bae}).
\end{remark}


\subsection{The quasi-classical limit of the BAE}

Let us expand the BAE (\ref{b-bae}) in the powers of $\eta$. Start with
\begin{equation*}
\begin{aligned}
\prod_{l=1}^{\l}\frac{1}{(v_k-\eps_l+\eta/2)(v_k+\eps_l+\eta/2)}= &
\prod_{j=1}^{\l}\frac{1}{v_k^2-\eps_j^2}\left(1-\frac{\eta}{2}\sum_{l=1}^{\l}\left(\frac{1}{v_k-\eps_l}+\frac{1}{v_k+\eps_l}\right)\right)+o(\eta),\\
\prod_{l=1}^{\l}\frac{1}{(v_k-\eps_l-\eta/2)(v_k+\eps_l-\eta/2)}= &
\prod_{j=1}^{\l}\frac{1}{v_k^2-\eps_j^2}\left(1+\frac{\eta}{2}\sum_{l=1}^{\l}\left(\frac{1}{v_k-\eps_l}+\frac{1}{v_k+\eps_l}\right)\right)+o(\eta),
\end{aligned}
\end{equation*}
\begin{equation*}
\begin{aligned}
\prod_{i\neq k}^{\l}(v_k-v_i+\eta)(v_k+v_i+\eta)=&\prod_{j\neq k}^{\l}(v_k^2-v_j^2)\left(1+\eta\sum_{i\neq k}^{\l}\left(\frac{1}{v_k-v_i}+\frac{1}{v_k+v_i}\right)\right)+o(\eta),\\
\prod_{i\neq k}^{\l}(v_k-v_i-\eta)(v_k+v_i-\eta)=&\prod_{j\neq k}^{\l}(v_k^2-v_j^2)\left(1-\eta\sum_{i\neq k}^{\l}\left(\frac{1}{v_k-v_i}+\frac{1}{v_k+v_i}\right)\right)+o(\eta).
\end{aligned}
\end{equation*}
Thus,
\begin{equation*}
\begin{aligned}
&\prod_{l=1}^{\l}\frac{1}{(v_k-\eps_l+\eta/2)(v_k+\eps_l+\eta/2)}\prod_{i\neq k}^{\l}(v_k-v_i+\eta)(v_k+v_i+\eta)=\\
= &\prod_{l=1}^{\l}\frac{1}{v_k^2-\eps_l^2}\prod_{i\neq k}^{\l}(v_k^2-v_i^2)
\left(1+\eta\left(\sum_{j\neq k}^{\l}\frac{2v_k}{v_k^2-v_j^2}-\sum_{m=1}^{\l}\frac{v_k}{v_k^2-\eps_m^2}\right)\right)+o(\eta),
\end{aligned}
\end{equation*}
\begin{equation*}
\begin{aligned}
&\prod_{l=1}^{\l}\frac{1}{(v_k-\eps_l-\eta/2)(v_k+\eps_l-\eta/2)}\prod_{i\neq k}^{\l}(v_k-v_i-\eta)(v_k+v_i-\eta)=\\
= &\prod_{l=1}^{\l}\frac{1}{v_k^2-\eps_l^2}\prod_{i\neq k}^{\l}(v_k^2-v_i^2)
\left(1-\eta\left(\sum_{j\neq k}^{\l}\frac{2v_k}{v_k^2-v_j^2}-\sum_{m=1}^{\l}\frac{v_k}{v_k^2-\eps_m^2}\right)\right)+o(\eta).
\end{aligned}
\end{equation*}
One can check that the first order contribution (i.e. first order in powers of $\eta$) from the
third term in the sum on the left hand side of (\ref{b-bae}) is zero:
\begin{equation*}
\begin{aligned}
&\left((\psi^-\phi^++\phi^-\psi^+)+2-2\sqrt{\psi^-\phi^-+1}\sqrt{\psi^+\phi^++1}\right)=\\
=&\left(2\psi\phi+\eta(\lambda\psi+\delta\phi)+\eta(\mu\psi+\gamma\phi)\right)+2-2(\psi\phi+1)-
2(\psi\phi+1)\frac{\eta}{2}\frac{(\lambda+\mu)\psi+(\delta+\gamma)\phi}{\psi\phi+1}+o(\eta)=o(\eta).
\end{aligned}
\end{equation*}
The first order contribution from the other two terms in (\ref{b-bae}) also gives zero:
\begin{equation*}
\begin{aligned}
&\frac{2\eta}{v_k}\left(v_k\sqrt{\psi\phi+1}-\xi\right)\left(v_k\sqrt{\psi\phi+1}+\xi\right)\prod_{l=1}^{\l}\frac{1}{v_k^2-\eps_l^2}\prod_{i\neq k}^{\l}(v_k^2-v_i^2)-\\
-&\frac{2\eta}{v_k}\left(v_k\sqrt{\psi\phi+1}+\xi\right)\left(v_k\sqrt{\psi\phi+1}-\xi\right)\prod_{l=1}^{\l}\frac{1}{v_k^2-\eps_l^2}\prod_{i\neq k}^{\l}(v_k^2-v_i^2)=0.
\end{aligned}
\end{equation*}
Thus we have to expand the BAE up to the second order. Start with
\begin{equation*}
\begin{aligned}
\sqrt{\psi^-\phi^-+1}=& \sqrt{\psi\phi+1}
\left(1+\frac{1}{2}\frac{\eta(\mu\psi+\delta\phi)+\eta^2\mu\delta}{\psi\phi+1}-\frac{1}{8}\frac{\eta^2(\mu\psi+\delta\phi)^2}{(\psi\phi+1)^2}\right)+o(\eta^2),\\
\sqrt{\psi^+\phi^++1}= & \sqrt{\psi\phi+1}
\left(1+\frac{1}{2}\frac{\eta(\lambda\psi+\gamma\phi)+\eta^2\lambda\gamma}{\psi\phi+1}-\frac{1}{8}\frac{\eta^2(\lambda\psi+\gamma\phi)^2}{(\psi\phi+1)^2}\right)+o(\eta^2).
\end{aligned}
\end{equation*}
Using this, we now calculate the $\eta^2$ contribution from the third term of (\ref{b-bae}):
\begin{equation*}
\begin{aligned}
&\delta\lambda+\mu\gamma-2(\psi\phi+1)\left[\frac{1}{4}\frac{(\mu\psi+\delta\phi)(\lambda\psi+\gamma\phi)}{(\psi\phi+1)^2}+\frac{1}{2}\frac{\mu\delta}{\psi\phi+1}
-\frac{1}{8}\frac{(\mu\psi+\delta\phi)^2}{(\psi\phi+1)^2}+\frac{1}{2}\frac{\lambda\gamma}{\psi\phi+1}-\frac{1}{8}\frac{(\lambda\psi+\gamma\phi)^2}{(\psi\phi+1)^2}\right]=\\
=&\,\frac{\left((\lambda-\mu)\psi+(\gamma-\delta)\phi\right)^2}{4(\psi\phi+1)}-(\gamma-\delta)(\lambda-\mu).
\end{aligned}
\end{equation*}
Using similar techniques, the first term of (\ref{b-bae}) gives the $\eta^2$ contribution
\begin{equation*}
\begin{aligned}
&\frac{2}{v_k}\prod_{l=1}^{\l}\frac{1}{v_k^2-\eps_l^2}\prod_{i\neq k}^{\l}(v_k^2-v_i^2)\Bigg[\left(v_k\sqrt{\psi\phi+1}-\xi\right)
\left(\frac{v_k}{2}\frac{\lambda\psi+\gamma\phi}{\sqrt{\psi\phi+1}}+\alpha+\frac{\sqrt{\psi\phi+1}}{2}\right)+\\
&+\left(v_k\sqrt{\psi\phi+1}+\xi\right)\left(\frac{v_k}{2}\frac{\mu\psi+\delta\phi}{\sqrt{\psi\phi+1}}+\beta+\frac{\sqrt{\psi\phi+1}}{2}\right)
+\left(v_k^2(\psi\phi+1)-\xi^2\right)\left(\sum_{j\neq
k}^{\l}\frac{2v_k}{v_k^2-v_j^2}-\sum_{m=1}^{\l}\frac{v_k}{v_k^2-\eps_m^2}\right)\Bigg],
\end{aligned}
\end{equation*}
and the second term gives the $\eta^2$ contribution
\begin{equation*}
\begin{aligned}
&\frac{2}{v_k}\prod_{l=1}^{\l}\frac{1}{v_k^2-\eps_l^2}\prod_{i\neq k}^{\l}(v_k^2-v_i^2)\Bigg[\left(v_k\sqrt{\psi\phi+1}+\xi\right)
\left(\frac{v_k}{2}\frac{\lambda\psi+\gamma\phi}{\sqrt{\psi\phi+1}}-\alpha-\frac{\sqrt{\psi\phi+1}}{2}\right)+\\
&+\left(v_k\sqrt{\psi\phi+1}-\xi\right)\left(\frac{v_k}{2}\frac{\mu\psi+\delta\phi}{\sqrt{\psi\phi+1}}-\beta-\frac{\sqrt{\psi\phi+1}}{2}\right)
-\left(v_k^2(\psi\phi+1)-\xi^2\right)\left(\sum_{j\neq k}^{\l}\frac{2v_k}{v_k^2-v_j^2}-\sum_{m=1}^{\l}\frac{v_k}{v_k^2-\eps_m^2}\right)\Bigg].
\end{aligned}
\end{equation*}
Summing up all terms we obtain
\begin{equation*}
\begin{aligned}
&2\prod_{l=1}^{\l}\frac{1}{v_k^2-\eps_l^2}\prod_{i\neq k}^{\l}(v_k^2-v_i^2)\Bigg[2(\alpha+\beta)\sqrt{\psi\phi+1}+2(\psi\phi+1)
-\frac{\xi\left((\lambda-\mu)\psi+(\gamma-\delta)\phi\right)}{\sqrt{\psi\phi+1}}+\\
&+2\left(v_k^2(\psi\phi+1)-\xi^2\right)\left(\sum_{j\neq k}^{\l}\frac{2}{v_k^2-v_j^2}-\sum_{m=1}^{\l}\frac{1}{v_k^2-\eps_m^2}\right)\Bigg]
+\frac{\left((\lambda-\mu)\psi+(\gamma-\delta)\phi\right)^2}{4(\psi\phi+1)}-(\gamma-\delta)(\lambda-\mu)=0.
\end{aligned}
\end{equation*}
Thus, we obtain the following BAE in the quasi-classical limit (keeping in mind the parameters
$\phi$, $\psi$, $\alpha$, $\beta$, $\gamma$, $\delta$, $\lambda$ and $\mu$ are defined in
(\ref{eta})):
\begin{equation}\label{qcbae}
\begin{aligned}
&(\alpha+\beta)\sqrt{\psi\phi+1}+(\psi\phi+1)-\frac{\xi\left((\lambda-\mu)\psi+(\gamma-\delta)\phi\right)}{2\sqrt{\psi\phi+1}}+\\
&+\left(v_k^2(\psi\phi+1)-\xi^2\right)\left(\sum_{i\neq k}^{\l}\frac{2}{v_k^2-v_i^2}-\sum_{l=1}^{\l}\frac{1}{v_k^2-\eps_l^2}\right)=\\
=&\frac{1}{4}\left((\gamma-\delta)(\lambda-\mu)-\frac{\left((\lambda-\mu)\psi+(\gamma-\delta)\phi\right)^2}{4(\psi\phi+1)}\right)
\frac{\prod_{l=1}^{\l}(v_k^2-\eps_l^2)}{\prod_{i\neq k}^{\l}(v_k^2-v_i^2)}.
\end{aligned}
\end{equation}
%
By setting $\beta=\psi=\phi=\delta=\mu=\xi=0$ in (\ref{qcbae}) we deduce the Bethe roots $\{v_k:k=1,...,\l\}$ appearing in (\ref{ev-final}) satisfy the BAE  
%
%
\begin{equation}\label{bae-final}
\frac{\alpha+1}{v_k^2}+\sum_{i\neq k}^{\l}\frac{2}{v_k^2-v_i^2}-\sum_{l=1}^{\l}\frac{1}{v_k^2-\eps_l^2}
=\frac{\gamma\lambda}{4v_k^2}\frac{\prod_{l=1}^{\l}(v_k^2-\eps_l^2)}{\prod_{i\neq k}^{\l}(v_k^2-v_i^2)}.
\end{equation}
\begin{remark}
Previous studies have used a correspondence between BAE and differential equations, through a generalised Heine-Stieltjes problem, as a route to numerically solve BAE for a wide range of models \cite{lmm15,zlm03,guan12,ml12,ld13,guan14,lm15}. 
It is interesting to observe that the BAE (\ref{bae-final}) have an interpretation as an {\rm inhomogeneous}, generalised Heine-Stieltjes problem. Define the polynomials
\begin{align*}
Q(x)&=\prod_{j=1}^{\l}(x-v_j^2), \\
P(x)&=\prod_{j=1}^{\l}(x-\eps_j^2).
\end{align*}
The BAE (\ref{bae-final}) are equivalent to the condition
\begin{align*}
v_k^2 P(v_k^2)Q''(v_k^2)+\left((\alpha+1)P(v_k^2)-v_k^2P'(v_k^2)\right)Q'(v_k^2) =\frac{\gamma\lambda}{4}\left[P(v_k^2)\right]^2.
\end{align*}
It follows that $Q(x)$ satisfies an inhomogeneous, linear, second-order differential equation
\begin{align*}
x P(x)Q''(x)+\left((\alpha+1)P(x)-xP'(x)\right)Q'(x) +V(x)Q(x)=\frac{\gamma\lambda}{4}\left[P(x))\right]^2
\end{align*}
where $V(x)$ is a Van Vleck polynomial of order ${\l}$. A similar correspondence also applies at the level of the BAE (\ref{qcbae}). 
\end{remark}

\subsection{Eigenvalues of the Hamiltonian}

Recall $\lambda_j^*$, given in (\ref{ev-final}), is the eigenvalue of the conserved
operator $\tau_j^*$ given in (\ref{tau-final}). 
To compute the eigenvalue of the Hamiltonian (\ref{hamiltonian}) consider
\begin{align*}
\sum_{j=1}^{\l}\eps_j^{-2}\lambda_j^*=&
\sum_{j,k:k\neq j}^{\l}\frac{1}{\eps_j^2-\eps_k^2}-\sum_{i=1}^{\l}\sum_{j=1}^{\l}\frac{2}{\eps_j^2-v_i^2}-\alpha\sum_{j=1}^{\l}\eps_j^{-2}=  \\
=&\sum_{i,j=1}^{\l}\frac{2}{v_i^2-\eps_j^2}-\alpha\sum_{j=1}^{\l}\eps_j^{-2}.
\end{align*}
From the BAE (\ref{bae-final}) we find 
\begin{equation*}
\sum_{j=1}^{\l}\frac{1}{v_i^2-\eps_j^2}=\frac{\alpha+1}{v_i^2}+\sum_{k\neq i}^{\l}\frac{2}{v_i^2-v_k^2}
-\frac{\gamma\lambda}{4v_i^2}\frac{\prod_{j=1}^{\l}(v_i^2-\eps_j^2)}{\prod_{k\neq i}^{\l}(v_i^2-v_k^2)}.
\end{equation*}
Thus,
\begin{equation*}
\begin{aligned}
\sum_{i=1}^{\l}\sum_{j=1}^{\l}\frac{2}{v_i^2-\eps_j^2}=&\,2(\alpha+1)\sum_{i=1}^{\l}{v_i^{-2}}+\sum_{i,k:i\neq k}^{\l}\frac{4}{v_i^2-v_k^2}
-\frac{\gamma\lambda}{2}\sum_{i=1}^{\l}\frac{1}{v_i^2}\frac{\prod_{j=1}^{\l}(v_i^2-\eps_j^2)}{\prod_{k\neq i}^{\l}(v_i^2-v_k^2)}=\\
=&\,2(\alpha+1)\sum_{i=1}^{\l}{v_i^{-2}}-\frac{\gamma\lambda}{2}\sum_{i=1}^{\l}\frac{1}{v_i^2}\frac{\prod_{j=1}^{\l}(v_i^2-\eps_j^2)}{\prod_{k\neq i}^{\l}(v_i^2-v_k^2)}.
\end{aligned}
\end{equation*}
This leads to
\begin{equation*}
\sum_{j=1}^{\l}\eps_j^{-2}\lambda_j^*=2(\alpha+1)\sum_{i=1}^{\l}v_i^{-2}
-\frac{\gamma\lambda}{2}\sum_{i=1}^{\l}\frac{1}{v_i^2}\frac{\prod_{j=1}^{\l}(v_i^2-\eps_j^2)}{\prod_{k\neq i}^{\l}(v_i^2-v_k^2)}-\alpha\sum_{j=1}^{\l}\eps_j^{-2}.
\end{equation*}
Implementing the change of variables $z_j=\eps_j^{-1},\ \ y_i=v_i^{-2}$ and setting $\lambda=-\gamma$ we obtain the expression
\begin{equation}\label{energy}
E=(1+G)\sum_{i=1}^{\l}y_i-\frac{1}{2}\sum_{k=1}^{\l}z_k^2
+\frac{\Gamma^2}{G}\sum_{i=1}^{\l}\frac{\prod_{j=1}^{\l}(1-y_iz_j^{-2})}{\prod_{k\neq i}^{\l}(1-y_iy_k^{-1})}
\end{equation}
for the eigenvalues of the Hamiltonian (\ref{H}) subject to the BAE obtained from
(\ref{bae-final}): 
\begin{equation}\label{bae-final-final}
1+G^{-1}+\sum_{i\neq k}^{\l}\frac{2 y_i}{y_i-y_k}+\sum_{l=1}^{\l}\frac{z_l^2}{y_k-z_l^{2}}
=-\frac{\Gamma^2}{G^2y_k}\frac{\prod_{l=1}^{\l}(1-y_k z_l^{-2})}{\prod_{i\neq k}^{\l}(1-y_ky_i^{-1})}.
\end{equation}



\section{Conclusion}
We have shown that the Hamiltonian (\ref{env}), describing a $p+ip$ pairing model interacting with its environment, is an integrable model. By mapping the Hamiltonian to the spin operator formalism (\ref{H}), we found  through a derivation of the model by the Boundary Quantum Inverse Scattering Method  that the energies of (\ref{H}) are given by (\ref{energy}) subject to the Bethe Ansatz Equations (\ref{bae-final-final}), and the operators (\ref{tau-final}) are conserved with eigenvalues given by (\ref{ev-final}). It is anticipated that this exact result will allow for a detailed analysis of the model in future studies. 

There are two examples in the literature \cite{dilz11,lrdo11} of related pairing models interacting with a single bosonic degree of freedom, where the boson-fermion interaction has the form    
\begin{align*}
\frac{\Gamma}{2}\sum_{\k}\left((k_x+ik_y)c_\k^\dagger c_{-\k}^\dagger b+(k_x-ik_y)c_{-\k}c_\k b^\dagger\right).
\end{align*}
This is in analogy with the system-environment interaction incorporated into (\ref{env}), however as mentioned in the Introduction for these models density matrices of the pairing model generically exhibit entanglement with the bosonic degree of freedom. These two models have vastly different ground-state behaviour, with \cite{lrdo11} exhibiting features which are qualitatively similar to those of the $p+ip$ pairing Hamiltonian while those of \cite{dilz11} are not. It will be interesting to see what (\ref{H}) has to offer in this regard, and what the consequences are of not accommodating entanglement with the environment. Another important line of future research is to extend the body of results on exact form factors and correlation functions for the $p+ip$ pairing model \cite{dilsz10,cdvv15}, and the same model interacting with a bosonic mode\cite{dilz11,cdvv}, to the integrable generalisation (\ref{env}).

\section*{Acknowledgements}
This research was supported by the Australian Research Council through Discovery Project
DP150101294. Inna Lukyanenko is funded through an International Postgraduate Scholarship
and a UQ International Scholarship.

\end{document}